\documentclass[twocolumn,preprintnumbers,superscriptaddress,nofootinbib,aps,prl,floatfix]{revtex4}

\usepackage{amsmath,amssymb}
\usepackage{dsfont}
\usepackage{graphicx} %loads the graphicx.sty package 
\usepackage{epstopdf} %loads the epstopdf.sty package 
\usepackage{slashed}
\usepackage{subfigure}
\usepackage{color}
\usepackage{multirow}

\usepackage{footmisc}

\usepackage{array}
\newcolumntype{L}[1]{>{\raggedright\let\newline\\\arraybackslash\hspace{0pt}}m{#1}}
\newcolumntype{C}[1]{>{\centering\let\newline\\\arraybackslash\hspace{0pt}}m{#1}}
\newcolumntype{R}[1]{>{\raggedleft\let\newline\\\arraybackslash\hspace{0pt}}m{#1}}

\newcommand{\be}{\begin{eqnarray*}}
\newcommand{\ee}{\end{eqnarray*}}

\newcommand{\bee}{\begin{eqnarray}}
\newcommand{\eee}{\end{eqnarray}}
\newcommand{\beeq}{\begin{equation}}
\newcommand{\eeeq}{\end{equation}}

\usepackage{xspace}

\begin{document}

\title{Cosmic ray air showers from sphalerons}

\begin{abstract}
The discovery of the Higgs boson marks a key ingredient to establish
the electroweak structure of the Standard Model. Its non-abelian gauge
structure gives rise to, yet unobserved, non-perturbative baryon and
lepton number violating processes. We propose to use cosmic ray air
showers, as measured at the Pierre Auger Observatory, to set a limit
on the hadronic production cross section of sphalerons. We identify
several observables to discriminate between sphaleron and QCD induced
air showers.
\end{abstract}
\author{Gustaaf Brooijmans} %\email{gusbroo@nevis.columbia.edu}
\affiliation{Physics Department, Columbia University, New York, NY 10027, United States of America\\[0.1cm]}
\author{Peter Schichtel} %\email{peter.schichtel@durham.ac.uk}
\affiliation{Institute for Particle Physics Phenomenology, Department
  of Physics,\\Durham University, DH1 3LE, United Kingdom\\[0.1cm]}
\author{Michael Spannowsky} %\email{michael.spannowsky@durham.ac.uk}
\affiliation{Institute for Particle Physics Phenomenology, Department
  of Physics,\\Durham University, DH1 3LE, United Kingdom\\[0.1cm]}

\preprint{IPPP/16/09,  DCPT/16/18, MCnet-16-02}
\maketitle

\section{Introduction}
\label{sec:intro}

The recent discovery of the Higgs boson~\cite{Aad:2012tfa,
  Chatrchyan:2012xdj} was the last missing piece to establish the
Standard Model of particle physics as effective theory describing
interactions at $\mathcal{O}(1)$~TeV. The Standard Model is predicted
to give rise to non-perturbative solutions at energies of
$\mathcal{O}(\alpha_W^{-1} m_W)~\mathrm{TeV} \simeq
\mathcal{O}(10)~\mathrm{TeV}$ which can result in the production of
many quarks, leptons and electroweak gauge bosons. This production of
multiple electroweak gauge bosons can occur with
\cite{Ringwald:1989ee, Espinosa:1989qn, McLerran:1989ab} or without
\cite{Cornwall:1990hh, Goldberg:1990qk, Goldberg:1991bc} baryon and
lepton number violating (BLNV) processes. The latter can be indicative
for the existence of electroweak sphalerons~\cite{Klinkhamer:1984di}, unstable solutions of the
classical action of motion for the Standard Model's $SU(2)_L$ that are
interpolating between topologically distinct vacua. Their discovery
would yield direct implications for the observed matter-anti-matter
asymmetry of the universe~\cite{Kuzmin:1985mm, Rubakov:1996vz,Cohen:1993nk}. However, whether these processes
can be observed at the LHC or a future collider remains an open
question as their production cross section is largely theoretically
unknown~\cite{Khoze:1990bm, Khoze:1991mx, Gibbs:1994cw, Ringwald:2003ns, Bezrukov:2003er, Tye:2015tva}.

Phenomenologically lepton-number violating processes with many gauge
bosons would give striking signatures at hadron colliders, easily
distinguishable from Standard Model backgrounds generated in
perturbatively describable interactions: events with many leptons,
missing energy and large $H_T$ are expected
\cite{Ringwald:2002sw, Ellis:2016ast}. Thus, the limiting factor to study
non-perturbative solutions of the Standard Model gauge group is the
center-of-mass energy of the initial state particles and the sphaleron
production cross section. While the LHC with up to $\sqrt{s} = 14$ TeV
is unlikely to be able to induce these processes, a future
proton-proton collider with $\sqrt{s} \simeq 100$ TeV might be able to
\cite{talkKhozeRingwald}.

Intriguingly, ultra high energy cosmic rays (UHECRs) provide us with a
natural source for proton-nucleon collisions, where the most energetic
ones, $E=10^{11}$ GeV, reach collision energies with nucleons in the
atmosphere of $\sqrt{s}\simeq\sqrt{2 m_N E} \simeq 500$ TeV. In this
paper we study whether the striking signatures of BLNV processes
induced by sphalerons can be observed at ground-based
detection experiments, e.g. the Pierre Auger Observatories.

Previous work aimed at setting limits on new physics using cosmic ray
interactions has either predominantly focused on exploiting 
primary and secondary neutrinos~\cite{Morris:1993wg, Ringwald:2001vk, Fodor:2003bn,Illana:2004qc} or hadronic shower particles~\cite{Illana:2006xg}. We
instead propose to study the longitudinal shower profile of
electro-magnetic particles with the fluorescens detectors at
Auger. Furthermore, we show the intense imprint the hard process
leaves in the shower distributions of muons at ground level.

In the following section we first discuss our analysis framework and
the potential final states induced. We then compare these signatures
with existing data as measured by the Auger Observatory and derive
actual limits on the production cross section of sphaleron processes
in the second section of this paper. In the last section we extend the
analysis assuming more detailed shower data was accessible. Finally we
provide a summary of our findings.

\section{Elements of the Analysis}
\label{sec:ana}

Calculating processes involving multi-vector boson final states
accompanied by several quarks and leptons is a very difficult task in
proton-proton collisions. Not only because the phase space is very
complex, but also because in our case the final state is induced by a
non-perturbative hard process.

For the signal events we use HERBVI~\cite{Gibbs:1994cw, Gibbs:1995bt} as implemented
in HERWIG~\cite{Marchesini:1991ch}, specifically designed to generate
BLNV processes. The BLNV process we study induces a change in baryon
and lepton number $\Delta B = \Delta L = -3$ and is assumed to be
\begin{equation}
  q q \to 7 \bar{q} + 3 \bar{l} + n_V W/Z + n_H H,
\end{equation}
where the incoming quarks and one outgoing antiquark are of first
generation, and three outgoing antiquarks are of each of the second
and third generations.

While it is for sphaleron-induced processes not necessary to involve
electroweak bosons, it was suggested that production cross sections
are enhanced if many electroweak bosons $\mathcal{O}(1/\alpha_W)$ are
produced in association with the fermions~\cite{Arnold:1987zg,
  Akiba:1988ay, McLerran:1989ab, Espinosa:1989qn,
  Ringwald:1989ee}. Hence we select $n_V = 24 $ and $n_H = 0$ in our
simulation.

To compute observables for Auger we further process the events with
CORSIKA~\cite{heck1998corsika} version 4.7. As default interaction
models we chose QGSJET~\cite{kalmykov1997quark} and
GHEISHA~\cite{gheisha}. The QCD background we compute with HERWIG as
well. To make sure that HERWIG handles the signal and background
collisions correctly we let CORSIKA also simulate primary collisions
on its own and compare to our Herwig results. We find very good
agreement, for example in the spacial number distribution of
secondaries over all energies.

\section{Observables and Limits from Auger}
\label{sec:auger}

The probability to produce a sphaleron in proton-proton collisions
from high-energetic proton-cosmic rays is readily parametrised by
\begin{equation}
\mathcal{P}_{\mathrm{sphal}} = A\,\sigma_{\mathrm{sphal}}/\sigma_T,
\end{equation}
where $A=14.6$ is the average atomic mass of a nucleus of air
\cite{Illana:2006xg} and $\sigma_T$ is the total cross section of a
proton with the air. The numerical value for $\sigma_T$ for
center-of-mass collision energies $\sqrt{s}$ corresponding to EeV
primaries we quote from~\cite{Collaboration:2012wt} to be
$505\pm22\,(\text{stat})\,^{+28}_{-36}\,(\text{sys})$ mb.

The Auger Observatory is a ground-based cosmic ray detector. It uses a
surface detector array (SD) consisting of 1600 water Cherenkov
detectors covering an area of $3000~\mathrm{km}^2$ and a fluorescence
detector (FD) to study detailed properties of cosmic ray showers in
the atmosphere. The combination of SD and FD allows the sampling of
electrons, photons and muons at ground level and the measurement of
the longitudinal development of air showers
\cite{Abraham:2009pm,Abreu:2010aa}.

\begin{figure}[!b]
  \includegraphics[width=0.43\textwidth]{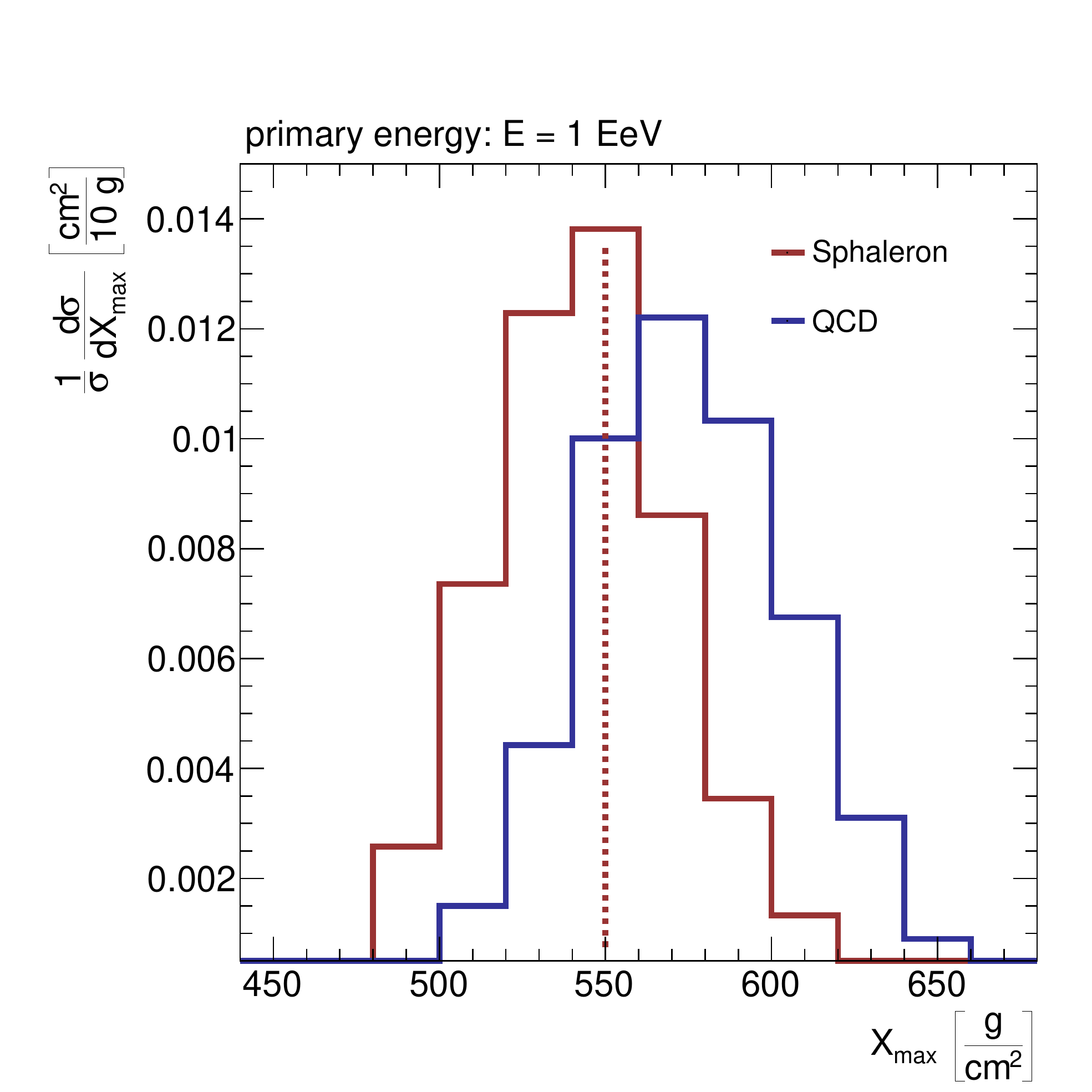}
  \caption{Distribution of $X_\text{max}$ for QCD and sphaleron
    induced events at $1$ EeV. We showered 2000 primary events for
    each sample.}
  \label{fig:xmax}
\end{figure}

The number distribution of particles in longitudonal direction can be
measured by the FD system. It follows the Gaisser-Hillas
function~\cite{gaisser1977reliability}. $X_\text{max}$ denotes the atmospheric depth, where
the number of electro-magnetic particles reaches its maximum. It can be used to measure the nature of cosmic rays~\cite{abraham2009upper}. In
Fig.~\ref{fig:xmax} we show the distribution of $X_\text{max}$ for QCD
and sphalerons at $E=1$ EeV. Both distributions approximate gaussians
of the same width, however, with clearly distinguishable mean
values. This is important information as the width can be used to
differ between protons and heavy
nuclei~\cite{deMelloNeto20141476}. Furthermore, we note that the mean
value depends not only on the short scale physics but also on the
collision energy, angle and interaction height. In Tab.~\ref{tab:xmax}
we show the expected $X_\text{max}$ for different inclinations and
heights for both QCD and sphaleron induced events. While the
dependence on the angle is rather strong this does not pose a problem
as the incident angle can be measured well by Auger\footnote{We use
  the vertical optical depth here and not the SLANT depth.}. The
dependence on the collision height only becomes significant when the
uncertainty on the primary interaction exceeds several kilometres. The
third column in each box indicates the background survival probability
$\epsilon_\text{B}$, after fixing the signal efficiency for a cut on
$X_\text{max}$ to $\epsilon_\text{S}=50\%$, see the dashed line in
Fig.~\ref{fig:xmax}.
\begin{table}[!b]
  \centering
  \renewcommand{\arraystretch}{1.2}
  \begin{tabular}{c|cccc}
    $ \text{log}\,E[eV] $& 17 & 18 & 19 & 20 \\ 
    \hline
    \hline
    \multirow{3}{*}{$20^\circ$ at 18.3 km}& $629 \pm 36$ & $696 \pm
    38$ & $757 \pm 41$ & $815 \pm 42$\\
    & $679 \pm 40$ & $739 \pm 42$ & $796 \pm 45$ & $849 \pm 47$\\
    & $0.10$ & $0.12$ & $0.22$ & $0.21$\\
    \hline
    \multirow{3}{*}{$45^\circ$ at 15.0 km}& $542 \pm 28$ & $594 \pm 
    29$ & $641 \pm 30$ & $682 \pm 33$\\
    & $580 \pm 31$ & $624 \pm 31$ & $672 \pm 34$ & $710 \pm 35$\\
    & $0.08$ & $0.15$ & $0.19$ & $0.23$\\
    \hline
    \multirow{3}{*}{$45^\circ$ at 18.3 km}& $491 \pm 28$ & $544 \pm 
    28$ & $590 \pm 31$ & $ 633 \pm 32$\\
    & $529 \pm 29$ & $576 \pm 32$ & $618 \pm 35$ & $660 \pm 35$\\
    & $0.07$ & $0.12$ & $0.20$ & $0.22$\\
    \hline
    \multirow{3}{*}{$45^\circ$ at 20.0 km}& $474 \pm 29$ & $525 \pm 
    28$ & $572 \pm 30$ & $616 \pm 33$\\
    & $513 \pm 31$ & $557 \pm 32$ & $600 \pm 35$ & $640 \pm 37$\\
    & $0.07$ & $0.12$ & $0.20$ & $0.21$\\
  \end{tabular}
  \caption{$\langle X_\text{max} \rangle$ for sphalerons (first
    number) and QCD (second number) as well as $\epsilon_\text{B}$ at
    $50\%$ signal efficiency $\epsilon_\text{S}$ for different primary
    energies, inclinations and interaction heights.}
  \label{tab:xmax}
\end{table}
The structure of $\langle X_\text{max} \rangle$ encourages us
to set a limit using a simple cut and count analysis. Asking for
$S/\sqrt{B}>2$ to set a $95\,\%$ confidence limit we can compute an
upper limit on the fraction of the total proton air cross section
\begin{align}
  f_{\sigma_T} \leq \sqrt{ \frac{ 4 \epsilon_\text{B} }{
      \epsilon_\text{S}^2 A^2 N }},
\end{align}
where $N$ is the number of recorded air showers. To estimate the
sensitivity of Auger we study its hybrid measurement mode. For the EeV
energy range~\cite{Unger:2007mc} found 4329 events recorded between
December 2004 and April 2007, while~\cite{abraham2009upper} found 3754
until 2009. Another anlysis~\cite{PierreAuger:2011aa} extracted 6744
events until 2011. We estimate that Auger has 10000 -- 15000 suitable
events by now. Cutting at $50\%$ signal efficiency yields
approximately a $10$--$20\%$ background efficiency. This yields a
limit of $(0.0008 - 0.0011) \times \sigma_T$,
e.g. $\sigma_\text{sphaleron} \le 500\, \mu$b. Furthermore, a
dedicated sphaleron analysis could also take lower energy data
($E\approx 10^{17}$ eV) into account, where the difference between
sphalerons and QCD is more pronounced. Additionally Auger may design
less stringent shower quality cuts streamlined for a dedicated
sphaleron analysis. Auger could therefore be in a position to limit
the sphaleron cross section to the level of few micro barn.

\section{Shower Observables for improved Limits}
\label{sec:dist}

So far we have shown that Auger is able to set an upper limit on the
sphaleron cross section using a simple cut and count approach for the
longitudinal shower profile. However, we expect a structural imprint
in each cosmic ray shower itself, which we can possibly connect to the
short distance physics during the collision. It is therefore our aim
in this section to identify additionally discriminating
observables in air showers. These will not be simply reconstructible
from the Auger detectors, but rather show the great power of cosmic
ray showers as window to new physics. We hope to trigger discussion in
the experimental community concerning the practical feasibility of
such measurements. In the following we only use showers with zero
inclination and fix the height of the primary collision to 18.3 km.

\begin{figure}[!b]
  \centering
  \includegraphics[width=0.43\textwidth]{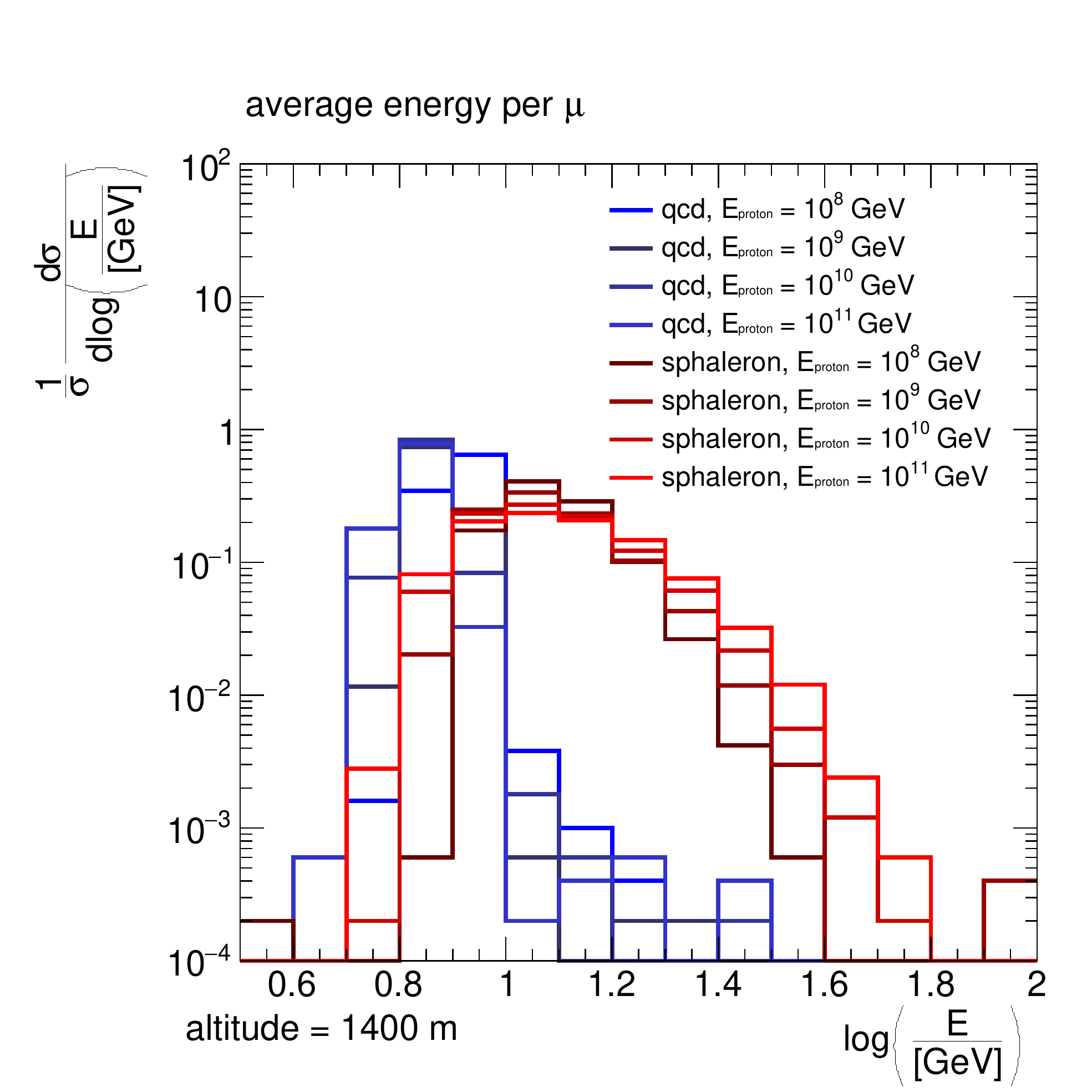}
  \caption{Expected average energy of a muon at the altitude of Auger
    for different primary energies. Sphalerons in red and QCD blue.}
  \label{fig:emean}
\end{figure}
\begin{figure}[!b]
  \includegraphics[width=0.43\textwidth]{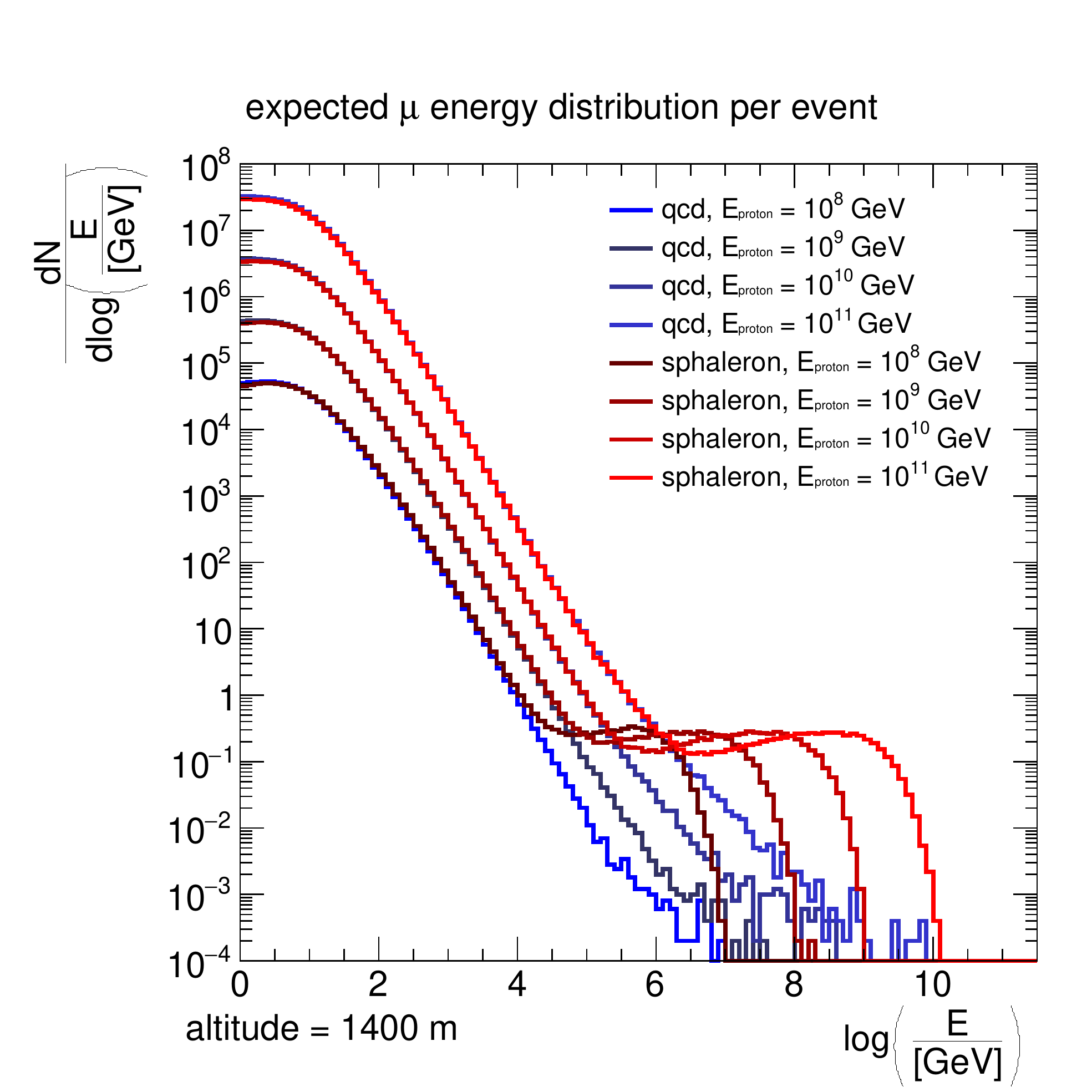}
  \caption{Expected energy distribution per event for muons.}
  \label{fig:edistr}
\end{figure}
\begin{figure}[!b]
  \includegraphics[width=0.43\textwidth]{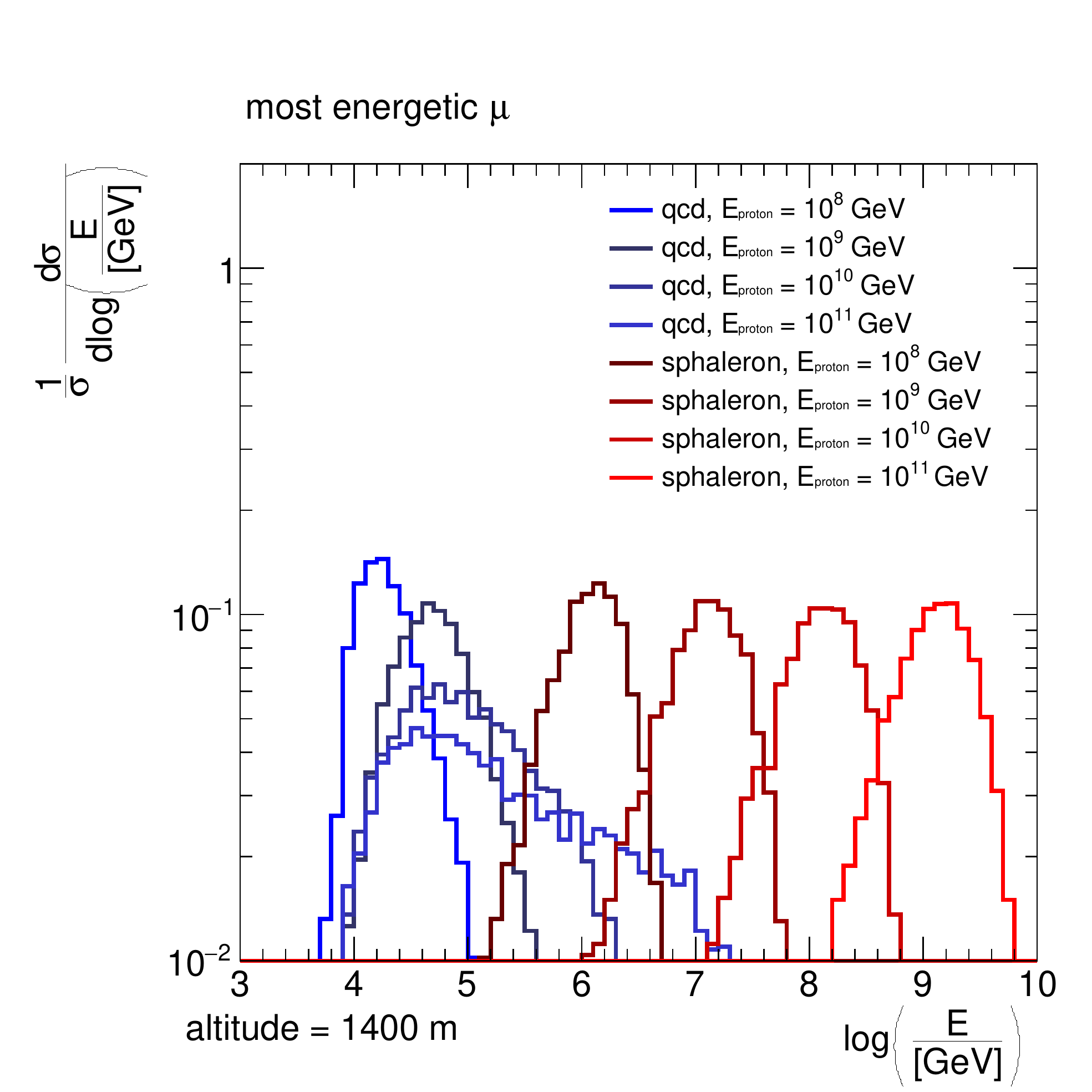}
  \caption{Distribution of the most energetic muon in sphaleron (red)
    and QCD (blue) events.}
  \label{fig:e1}
\end{figure}
\begin{figure}[!b]
  \includegraphics[width=0.43\textwidth]{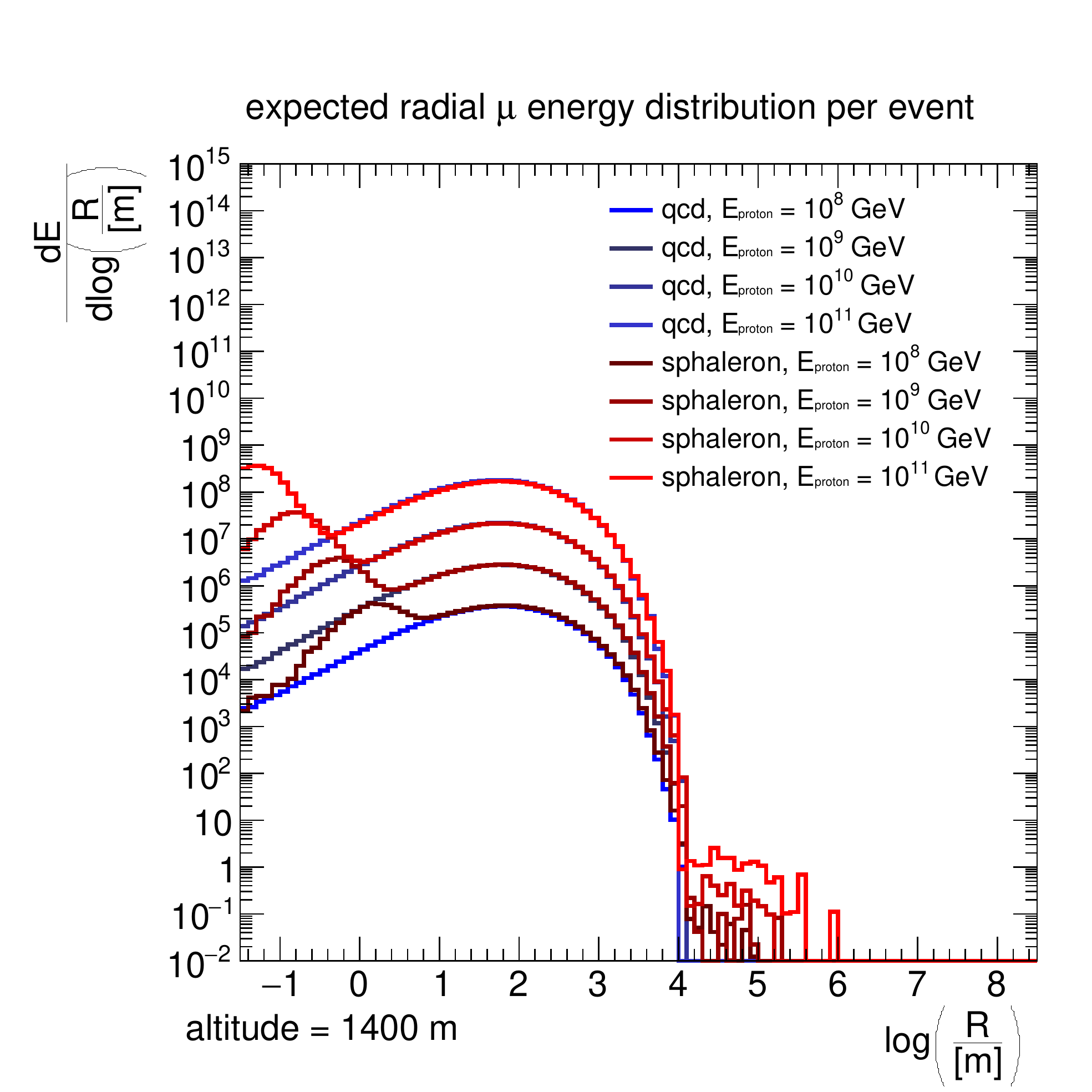}
  \caption{Radial energy distribution.}
  \label{fig:eradius}
\end{figure}

In Fig.~\ref{fig:emean} we plot the expected average energy a muon
carries when reaching the Cherenkov chambers. We observe a huge
difference between the sphaleron induced events and QCD. To trace back
this difference we first exploit the expected energy distribution per
event, see Fig.~\ref{fig:edistr}. QCD and sphalerons look exactly the
same except for the high energy region, where the sphaleron
distributions are enhanced. Indeed, as we show in Fig.~\ref{fig:e1},
it is almost exclusively the highest energy muon which induces this
difference. To learn more we also plot the radial energy distribution
in Fig.~\ref{fig:eradius}. Again QCD and sphaleron events almost agree
completely except for the shower core. We can therefore conclude that
a sphaleron event will most likely be accompanied by a very highly
energetic muon within its shower core. Tagging this one muon
constitutes a powerful method to observe sphaleron induced air
showers. In addition we note that sphaleron events are significantly
bigger than QCD events, see Fig.~\ref{fig:eradius}. While all energy of QCD events is confined in a radius of less than 10 km around the primary collision point, sphaleron events can induce air showers with radii of 100 km and more. Some of the quarks and gauge bosons in the sphaleron decay can have large transverse momenta. Although the flux of muons is small in the most outer part of the air shower, it can be used as a smoking gun signature for sphalerons.

While the former observables are difficult to measure, as they are relying on the
experiment's ability to measure the muons' energy, their structure is rather simple.
Let us now assume that we have a perfect detector layer on the ground,
able to measure the spatial and energetic distribution of all muons of the air shower. To search for structural differences we can cluster the
muons with a jet algorithm, a very promising strategy at
the LHC~\cite{Abdesselam:2010pt}. We use an anti-$k_T$ algorithm in
spherical coordinates as implemented in FastJet~\cite{Cacciari:2011ma}
with radius parameter $R=0.2$. This choice guarantees good performance
in the forward region as well as circular jet shapes when
projected onto a sphere. However, due to the practical necessity of including a 
thinning procedure during air shower evolution\footnote{We process several thousand events per run.}, we are
confronted with a structural problem: the thinning algorithm (over)simplifies the kinematics of soft muons. To avoid a strong sensitivity of our observables to badly modelled soft particles, we only allow muon clusters with $E>10$ GeV to be recombined into jets.
\begin{figure}[!b]
  \includegraphics[width=0.43\textwidth]{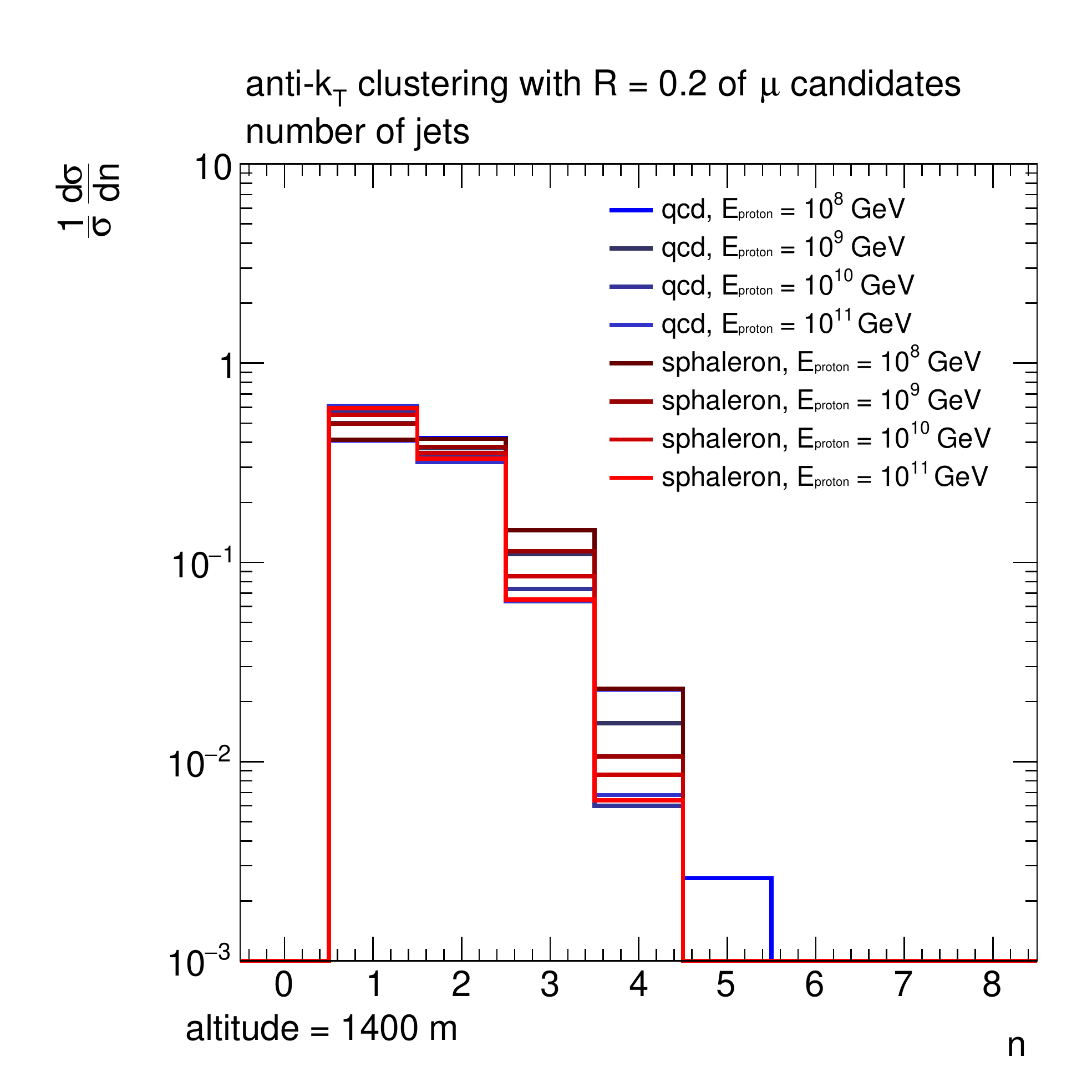}
  \caption{Number of jets from anti-$k_T$ $R=0.2$ jet clustering algorithm.}
  \label{fig:njets}
\end{figure}
\begin{figure}[!b]
  \includegraphics[width=0.43\textwidth]{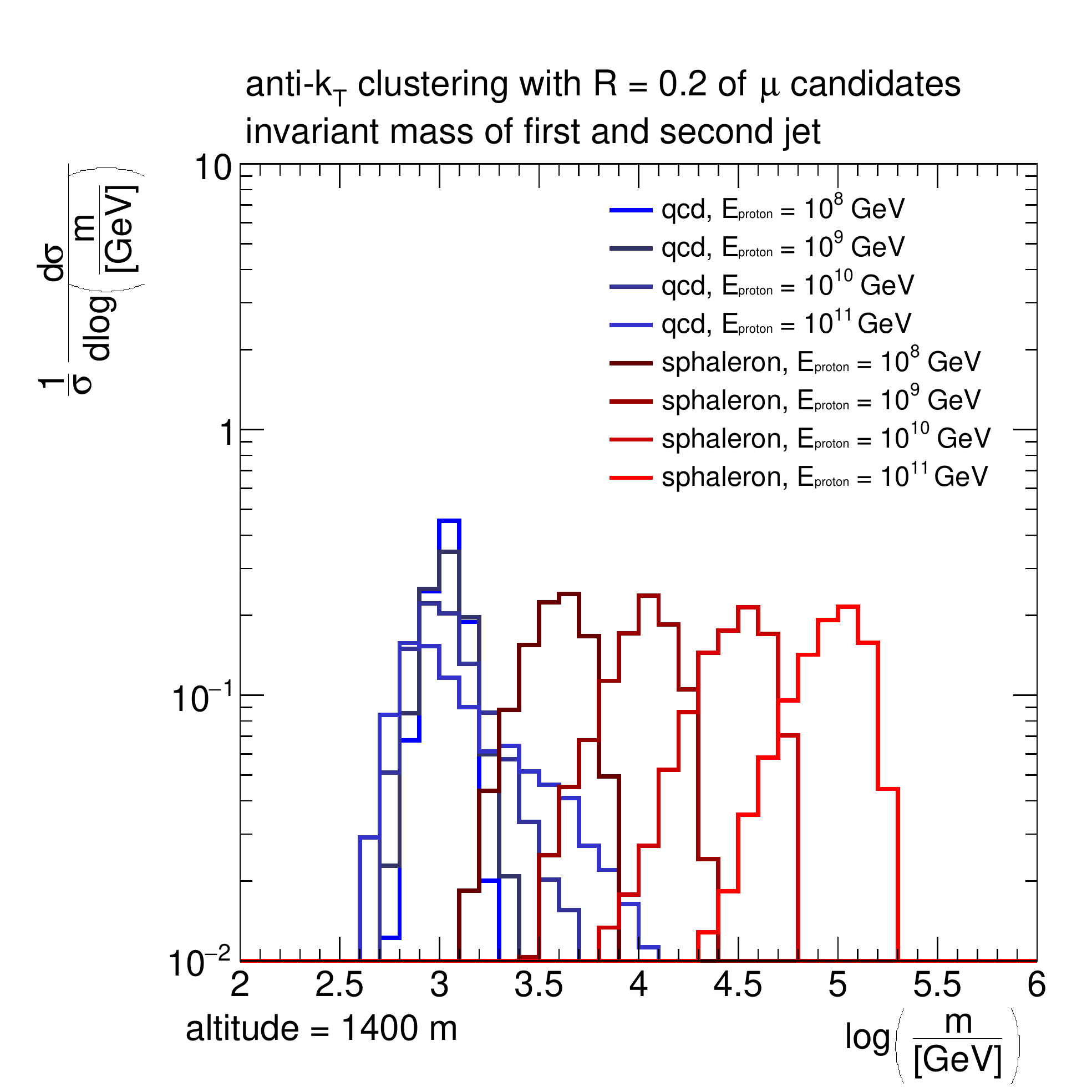}
  \caption{Invariant mass of the sum of the first and second leading
    jet in energy.}
  \label{fig:mass}
\end{figure}

In Fig~\ref{fig:njets} we show the number of muon-jets, an observable
with sensitivity to decay structure. However, we can observe no
difference between both hypothesis. The situation changes once we sum
over the two\footnote{Summing over more jets does not change the
  result any further.}  leading jets in energy and compute the
invariant mass, see Fig~\ref{fig:mass}. Obviously, there is a strong correlation between this observable and energy distribution of the hardest muon in Fig.~\ref{fig:e1}. 

\section{Conclusions and Outlook}
\label{sec:conclusion}

Using $X_\text{max}$, a well known air shower observable, we are able to set an
upper limit on the sphaleron cross section $\sigma_\text{sphaleron} \le 500\,\mu$b assuming an amount of 10000 -- 15000 events by now. We encourage the experimentalists to perform a dedicated sphaleron analysis which might set even more stringent limits of the order of several $\mu$b.

We introduce new air shower observables connected to the energy
distribution of the muons and show that sphaleron events most like
have one very hard muon in their shower core. Any experiment to measure
these hard muons could help to identify sphaleron events in air showers. A second observable we identified to provide a strong discrimination between sphaleron and QCD events is the radial size of the air shower. 

In the last part of the paper we use jet clustering, a technique well
developed in collider physics, on air shower muons. We demonstrate
that we can recover the powerfull discriminating behaviour observed
for the energy distribution before. However, from a technical point of
view the thinning algorithm poses an obstacle. Because the algorithm
is computer wise neccessary we propose an enhancemnt, where, for
example, the particles are not just dropped but clusterd to ghost
particles to have directional information available as well. This,
however, is clearly a field of further study.

We hope to initiate discussion in the community about the technological feasibility of such measurements. Even if one eventually concludes that the Auger Observatory will not be able to exploit these observable, it might be intriguing to use the wide and densely packed coverage of smart phones \cite{Whiteson:2014kca} to search for non-perturbative solutions of the Standard Model.
\\

{\emph{Acknowledgements.}}  

We thank Martin Erdmann, Valya Khoze and Ralf Ulrich for helpful
discussions. We thank Brian Webber for assistance with HERBVI. We
thank Tanguy Pierog for assistance with CORSIKA. P.\,S. acknowledges
support by the European Union as part of the FP7 Marie Curie Initial
Training Network MCnetITN (PITN-GA-2012-315877).

%

%%%%%%%%%%%%%%%%%%%%%%%%%%%%%%%%%%%%%%%%%%%%%%%%%%%%%%%%%%
\bibliography{references}

\end{document}